\documentstyle[psfig]{mn} 
\title{The intermediate-age open cluster 
 NGC~2112\thanks{Based on observations  
carried out at Mt Ekar, Asiago, Italy. All the photometry is available 
at WEBDA database: http://obswww.unige.ch/webda/navigation.html}} 
\author[Carraro  at al.]        
{Giovanni Carraro$^1$, Roberto Barbon$^2$, and Carla S. Boschetti$^1$
\thanks{email: 
giovanni.carraro@unipd.it (GC); barbon@pd.astro.it (RB);
boschetti@pd.astro.it(CB) }\\ 
$^1$Dipartimento di Astronomia, Universit\`a di Padova, Vicolo 
dell'Osservatorio 2, I-35122 Padova, Italy \\
$^2$Osservatorio Astrofisico di Asiago, Universit\'a di Padova, I-36012,
Asiago (VI), Italy\\
 } 
 
\date{\it Submitted: March 2002} 
\pubyear{2002} 
\begin{document} 
\maketitle 
\title{The open cluster NGC~2112} 
 
\begin{abstract} 
We report on $BVI$ CCD photometry of a field centered  
on the region of  
the intermediate-age open cluster NGC~2112 down to $V=21$. 
Due to the smaller field coverage, we are able to limit
the effect of field star contamination which hampered in the past 
precise determinations of the cluster age and distance.
This way, we provide updated estimates of NGC~2112 fundamental
parameters. Having extended the photometry to the $I$ pass-band,
we are able to construct  a colour-colour diagram, from which  
we  infer a reddening $E_{B-V}= 0.63\pm0.14$ mag.
The comparison of the Colour-Magnitude Diagram (CMD) with theoretical
isochrones leads  to 
a distance of $850 \pm 100$ pc, and an age 
of  $2.0 \pm 0.3$ Gyr. While the distance is in agreement
with previous determinations, the age turns out to be much better constrained
and significantly lower than previous estimates. 
\end{abstract} 
 
\begin{keywords} 
Open clusters and associations: general -- open clusters and associations:  
individual: NGC~2112 - Hertzsprung-Russell (HR) diagram 
\end{keywords}

\section{Introduction} 
NGC~2112 (Collinder~76, C~0551-0031, OCL~509)  
is a northern open cluster of intermediate-age, 
located relatively far from the Galactic plane toward 
the anti-center direction ($\alpha=05^{\rm h}~53^{\rm m}.9$,  
$\delta=+00^{\circ} 
23^{\prime}$, $l=205^{\circ}.91$, $b=-12^{\circ}.59$, J2000.).  
It is classified as a II2m open cluster by Trumpler 
(1930), and has a diameter of about $18^{\prime}$, according to
Lyng\aa~ (1987). 
It is quite a poorly studied object, but rather interesting
due  to its position in the disk and to its combination
of suspected old age and low metal abundance, which
would make it a noteworthy object to study in the framework
of the chemical evolution of the Galactic disk (Carraro et al. 1998). 
Moreover
it remained unstudied for long time due mainly, we guess, to the high
contamination of field stars toward its direction
which prevented precise estimates of its age and distance
insofar (Richtler \& Kaluzny 1989).\\
For these reasons, 
we decided to undertake a multicolour  
CCD study of the cluster, as presented 
in the present paper, which  
is the fourth of a series dedicated at improving 
the photometry of northern intermediate-age open clusters at Asiago 
Observatory. We already reported elsewhere on NGC~1245 
(Carraro \& Patat 1994), on NGC~7762 (Patat \& Carraro 1995) and
on NGC~2158 (Carraro et al.  2002). 
 
\noindent
The plan of the paper is as follows. In Sect.~2 we summarize 
the previous studies on NGC~2112, while Sect.~3 illustrates  
the observation and reduction strategies. 
The analysis of the CMD is performed in Sect.~4, whereas 
Sect.~5 deals with the determination of cluster reddening, 
distance and age. Sect.~6 is dedicated to discuss
the properties of the cluster in the context of the
Galactic disk chemical evolution. Sect.~7 is devoted
to analyze the geometrical
structure and star counts and, 
finally, Sect.~8 summarizes our findings. 
 
\begin{figure} 
\centerline{\psfig{file=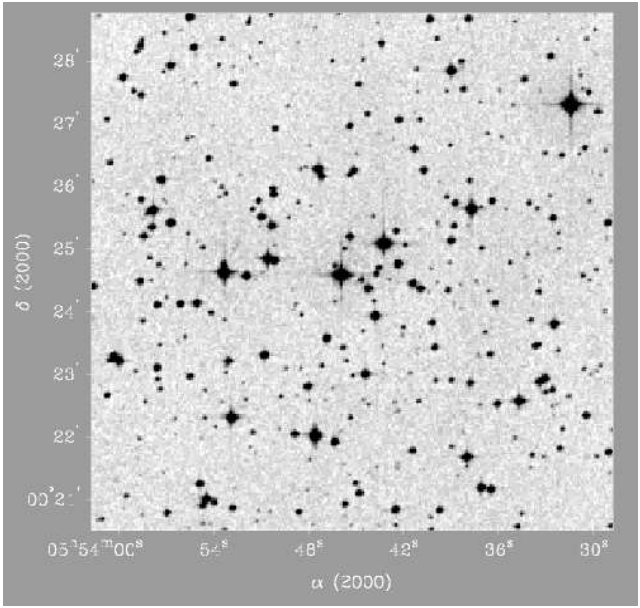,width=\columnwidth}} 
\caption{A DSS image of a region around NGC~2112
covered by the present study. North is up, east on
the left.}
\label{mappa} 
\end{figure} 
 
\section{Previous investigations} 
NGC~2112 has been already studied in the past. 
The first investigation was carried out by Richtler (1985), 
who obtained photographic $BV$ photometry for about 80 
stars down to $V=15$. Although he was not able to reach the cluster Turn
Off (TO), he  nevertheless drew the attention on this probably old
neglected cluster, and suggested that NGC 2112 had a reddening
of about 0.5 mag, and was about 800 pc far from the Sun.
By analyzing additional Str\"omgren photometry, Richtler
suggested that the cluster had to be very metal poor ($[Fe/H]
\approx -1$).\\
A more accurate and deeper analysis was performed by Richtler and
Kaluzny (1989). They obtained $BV$ CCD photometry for about
500 stars in a field of 200 squared arcmin. Additionally, they obtained
moderate resolution spectra for a handful of bright stars.
Their conclusions were that the cluster was very contaminated by field
stars. Nevertheless they were able to strengthen the suggestions of
Richtler (1985) by claiming that the cluster was 3-5 Gyrs old, 
700-800 pc far from the Sun, and had a reddening E$(B-V)~=~0.60$ mag.
As for the metal abundance, they could not  conclude anything, since they
were not able to separate cluster probable members from non-members.\\
The Washington photometry reported
by Geisler (1987) and Geisler et al. (1991) apparently corroborated 
the previous suggestions
that NGC~2112 was very metal poor ($[Fe/H] \approx -1.3$), 
but this finding hardly
reconciled with the age inferred by the previous photometries
in the conventional frame of the Galactic disk evolution
(Richtler \& Kaluzny 1989).
Only recently, however, 
the metal abundance of NGC~2112 was derived by means of high 
resolution spectroscopy by Brown et al. (1996).
They were able to select a few cluster members on the basis
of radial velocity, providing at the end a much higher metal content
value $[Fe/H]~=-0.15$, only slightly lower than the solar one.
These authors emphasized that the previous lower metallicity estimates
were due to the difficulty to separate cluster members and 
to the uncertainties
in the reddening estimate. The metallicity they obtained was however
based on only one certain member and two probable members.
Clearly, as the same authors stressed, an astrometric study
is urgent for this cluster in order to infer a more
reliable  abundance measurement.\\
\noindent
To summarize the present situation 
regarding the cluster basic parameters, we note that:\\
{\it (i)} as for the distance, Richtler (1985) and Richtler \& Kaluzny (1989)
report {\it preliminary} and {\it of the order of} estimates,
respectively;\\
{\it (ii)} according to Richtler (1985), the cluster is simply {\it very old},
whereas Richtler \& Kaluzny (1989) suggest an age between 3 and 5 Gyrs.\\
{\it (iii)} reddening determinations are reported as 
indications only;\\
{\it (iv)} as far as the metallicity is concerned, one might rely
on Brown et al. (1996) modern result, but we shall devote some
discussion to the possibility of this cluster to be also very metal poor.\\
In conclusion, NGC~2112 has the reputation to be a
severely contaminated open cluster
for which some of the fundamental parameters revealed 
very difficult to be estimated.
For these reasons we thought worthwhile to carry out a new photometric study of
this cluster.
 
\begin{table} 
\tabcolsep 0.20truecm 
\caption{Journal of the observations of NGC~2112 
and Landolt fields (November 14 , 2001).} 
\begin{tabular}{ccccc} 
\hline 
\multicolumn{1}{c}{Field}    & 
\multicolumn{1}{c}{Filter}    & 
\multicolumn{1}{c}{Integration time}& 
\multicolumn{1}{c}{Seeing}    &     
\multicolumn{1}{c}{Airmass}\\ 
      &        & (sec)     & ($\prime\prime$) &\\ 

\hline 
PG~1047+003 &     &                   &      & \\ 
            & $B$ &  60,60            &  2.0 & 1.440\\ 
            & $V$ &  30,30            &  2.1 & 1.450\\ 
            & $I$ &  30,30            &  1.8 & 1.438\\ 
 NGC~2112   &     &                   &      & \\ 
            & $B$ &  60,600           &  2.0 & 1.438\\ 
            & $V$ &  5,15,30,30,300   &  1.9 & 1.453\\ 
            & $I$ &  2,300            &  2.3 & 1.445\\
PG~2331+055 &     &                   &      & \\ 
            & $B$ &  120,120          &  2.0 & 1.308\\ 
            & $V$ &  30,30,60,60      &  2.1 & 1.307\\ 
            & $I$ &  30,30,30         &  1.9 & 1.307\\
 SA~93      &                         &      & \\ 
            & $B$ &  60,60            &  2.1 & 1.514\\ 
            & $V$ &  30,30,30         &  2.0 & 1.500\\ 
            & $I$ &  30,30            &  2.1 & 1.497\\
\hline 
\end{tabular} 
\end{table}

\begin{figure}  
\centerline{\psfig{file=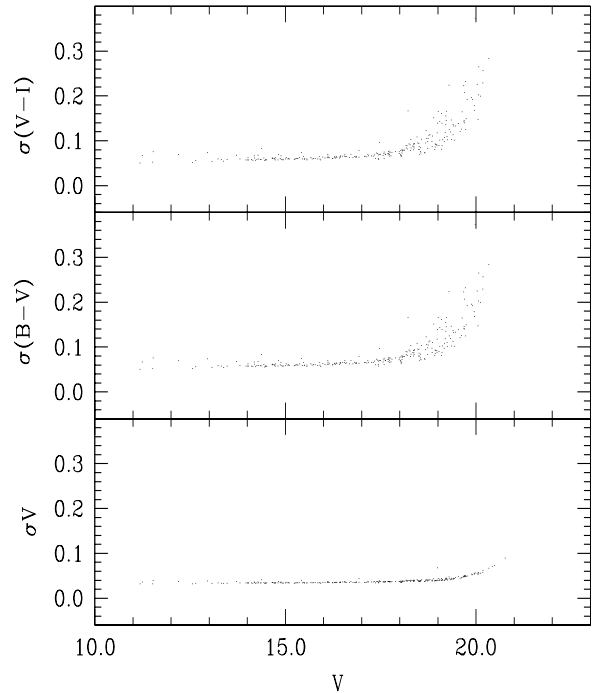,width=8cm,height=10cm}} 
\caption{Photometric errors as a function of magnitude, for our NGC~2112 
observations.} 
\label{fig_errors} 
\end{figure} 
 
\section{Observations and Data Reduction} 
 
Observations of NGC~2112 were carried out with the AFOSC camera at the 
1.82~m telescope of Cima Ekar (Asiago, Italy), in the night of November 14, 
2001. AFOSC samples a $8^\prime.14\times8^\prime.14$ field in a  
$1K\times 1K$ thinned CCD. The typical seeing was around 2.0  
arcsec. Additionally, we provide $V$ photometry of a field
10 arcmin eastward of the cluster to derive field star counts.\\

We also observed the Landolt (1982) standard star
fields PG~1047+003, SA~93 and PG~2331+055. The details of the observations
are listed in Table~1, where the observed fields, together
with exposure times, seeing and airmasses are reported.\\
A DSS\footnote{Digitiged Sky Survey,
http://www.eso.org/dss/dss} map of the observed region is shown in Fig.~1.
 
The data have been reduced by using the IRAF\footnote{IRAF  
is distributed by the National Optical Astronomy Observatories, 
which are operated by the Association of Universities for Research 
in Astronomy, Inc., under cooperative agreement with the National 
Science Foundation.} packages CCDRED, DAOPHOT, and PHOTCAL. 
The calibration equations obtained are: 

\[
b=B+1.407\pm0.022+(0.004\pm0.029)(B-V)+0.30\,X   
\]

\[ 
v=V+0.752\pm0.009+(0.036\pm0.012)(B-V)+0.18\,X   
\]

\[
i=I+1.619\pm0.048+(0.011\pm0.052)(V-I)+0.08\,X 
\]

\noindent
where $BVI$ are standard magnitudes, $bvi$ are the instrumental  
ones, and $X$ is the airmass. The standard stars in these fields
provided a very good colour coverage.
For the extinction coefficients, 
we assumed the typical values for the Asiago Observatory. 
The error affecting these coefficients
is 0.03 (Desidera et al. 2001).

\begin{figure*} 
\centerline{\psfig{file=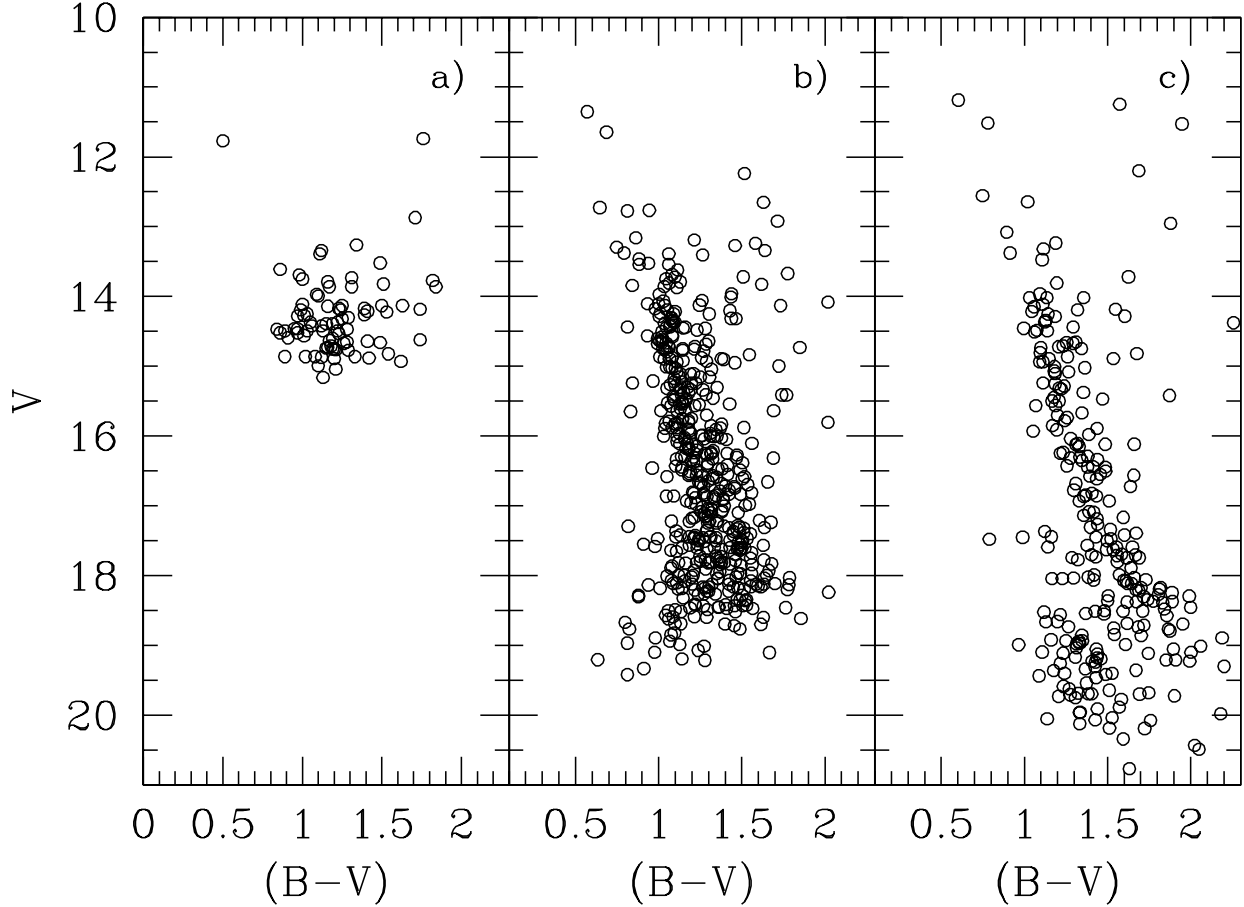,width=\textwidth}} 
\caption{$BV$ CMDs of NGC~2112. {\bf Panel {\it a)}}: Richtler 
(1985) photometry. {\bf Panel {\it b)}}: Richtler \& Kaluzny (1989) 
photometry. {\bf Panel {\it c)}}: the photometry presented in this study.}
\end{figure*} 
 
Finally, Fig.~2 presents the run of photometric errors 
as a function of magnitude. These errors take into account fitting errors 
from DAOPHOT and calibration errors, and have been computed 
following Patat \& Carraro (2001). 
It can be noticed that stars brighter than  
$V \approx 20$ mag have  
photometric errors lower than 0.1~mag in magnitude
and lower than 0.2~mag in colour. The final photometric data are  
available in electronic form at the  
WEBDA\footnote{http://obswww.unige.ch/webda/navigation.html} site. \\
We compared our photometry with the CCD one from 
Richtler \& Kaluzny (1989), finding a nice agreement.
In fact, from the 53 stars in common
brighter than $V \approx 16.5 $ mag, we obtain the
following mean differences:

\[
V_{CBB} - V_{RK} = -0.028\pm0.067   (s.e.)
\]
 
\[
(B-V)_{CBB} - (B-V)_{RK} = 0.034\pm0.047   (s.e.)
\]

\noindent
where the suffix $CBB$ refers to this study, whereas
the suffix $RK$ refers to the Richtler \& 
Kaluzny (1989) photometry.

\section{The Colour-Magnitude Diagrams} 
A comparison of our photometry with past analyses is shown in Fig.~3,
from which it is evident that the present study supersedes the 
previous ones. Richtler's (1985) photometry (panel {\it a}) 
does not reach the TO, whereas the photometry of Richtler \& Kaluzny
(1989), panel {\it b}, has:\\
{\it (i)} photometric errors amounting to
$\Delta V = 0.1$ and $\Delta (B-V) = 0.15$ already at $V \approx 18.0$
(see their Fig.~4a,b); \\
{\it (ii)} covers a very large region of the sky, enhancing this way
the effect of the field star contamination.\\
This is mostly evident by considering the width of the MS, which in
our case (panel {\it c}) is much narrower. 
By inspecting these CMDs, we find that the TO is located at 
$V \approx 14.5$ mag, $(B-V) \approx 1.0$ mag and the MS is clearly visible
down to V=19 mag. A binary sequence is evident on the right of the MS 
in both Richtler \& Kaluzny (1989) and  our photometry.
The evolved region of the CMD is poorly populated. This is not
surprising, since the cluster is not much populous. 
Probably this fact is the most effective in hampering a precise
age estimate. Nevertheless the sub-giant branch is sufficiently clear,
and probably there are 3-4 stars in the Red Giant Branch (RGB), among
which some of those
used by Brown et al. (1996) to estimate the metallicity of NGC~2112.
Apparently, no RGB clump stars are present.\\
All the blue stars above the cluster TO might be field stars located
between us and the cluster, or blue straggler stars, which we know
to be present in many intermediate-age open clusters
(Mermilliod 1982). 
The field star contamination is much evident in panel {\it b},
where the MS starts widening significantly already at V = 17 mag,
where the photometric errors are less than 0.1 mag (see above) 
and the TO region is much more contaminated by
interlopers.
In our photometry (panel {\it c}) the field contamination is less severe
due to the smaller field coverage. 
Clearly, the bulk of stars leftwards of the MS
below $V \approx 19.0$ mag, are field stars.
 
\begin{table*}  
\tabcolsep 0.7truecm 
\caption {NGC~2112 fundamental parameters taken from the literature.} 
\begin{tabular}{ccccc}  
\hline 
&Richtler (1985) & Richtler \& Kaluzny (1989) & Geisler (1987) & Brown et al. (1996) \\ 
\hline 
$E(B-V)$   &  0.50  &  0.60 & & 0.60  \\ 
$(m-M)$    &  9.2  & 9.2-9.4 &&   \\ 
distance (pc) & 700   & 700-800 & &  \\ 
Age (Gyr)  &  Very old  & 3-5  & &     \\
$[Fe/H]$   & -1.0 & -1.& -1.3 & -0.15\\  
\hline 
\end{tabular} 
\end{table*}

\section{Cluster fundamental parameters} 
As we said in Sect.~2, 
the fundamental parameters of NGC~2112 (age, distance,
metallicity  and reddening)
are still poorly known (see Table~2). 
The cluster age estimates range from  the simple statement
that the cluster is very old to 3-5 Gyr, 
the distance from 700 to 800~pc and the reddening E$(B-V)$ from 0.5 
to 0.6 mag. None of these 
previously reported estimates has an error bar and we take them as first
guesses only.  
In the next Sections we are going to derive updated estimates for 
them.

\subsection{Reddening} 
In order to obtain an estimate of the cluster mean reddening, we 
analyze the distribution of the MS stars with $V < 18$ in the $(B-I)$  
vs. $(B-V)$ plane, which is shown in Fig.~4. 
 
The linear fit to the main sequence in the $(B-I)$  
vs. $(B-V)$ plane, 
\begin{equation} 
(B-I) = Q + 2.25 \times (B-V) 
\end{equation} 
can be expressed in terms of $E_{B-V}$, for the $R_V=3.1$ 
extinction law, as 
\begin{equation} 
E_{B-V} = \frac{Q-0.014}{0.159} \,\,\,\,  , 
\label{mu2} 
\end{equation} 
following the method proposed by  Munari \& Carraro (1996). 
This method provides an estimate of the mean reddening and, 
as amply discussed in Munari \& Carraro (1996), can be used 
only for certain colour ranges. In particular Eq.~(\ref{mu2}) 
holds over the range $-0.23 \leq (B-V)_0 \leq +1.30$. 
MS stars have been selected by considering all the stars
having $15 \leq V \leq 18$ and  
$0.9 \leq (B-V) \leq 1.7$. 
A least-squares  fit through all these stars 
gives $Q=0.115$, 
which, inserted in Eq.~(\ref{mu2}), provides  $E_{B-V}=0.63\pm0.14$ mag. 
The uncertainty is rather large, and is due to the scatter 
of the stars in this plane, which indicates the presence 
of stars with different reddening, presumably a mixture 
of stars belonging to the cluster and to the field. 
Besides, also some differential reddening cannot be ruled out, 
as it was proposed in past photometries, due to the location of the cluster,
in the eastern part of Orion, engulfed in the Barnard's loop HII region
(SH2-276, Sharpless 1959).
 
\begin{figure} 
\centerline{\psfig{file=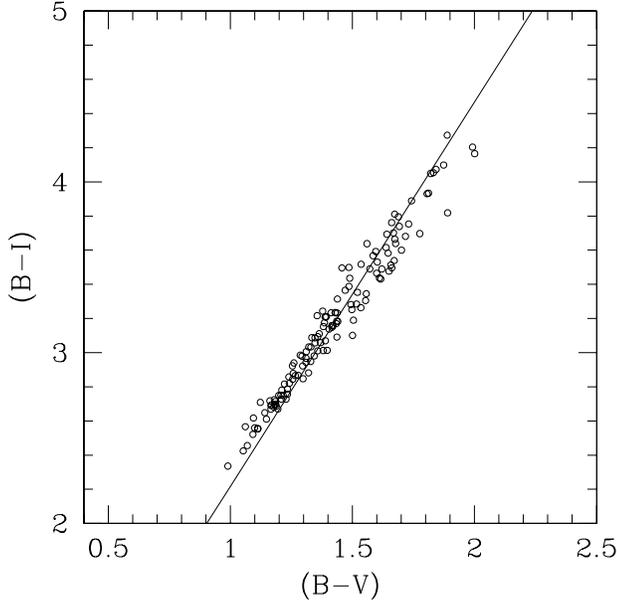,width=\columnwidth}} 
\caption{NGC~2112 MS stars 
in the $(B-V)$ vs. $(B-I)$ plane.} 
\end{figure}

\begin{figure} 
\centerline{\psfig{file=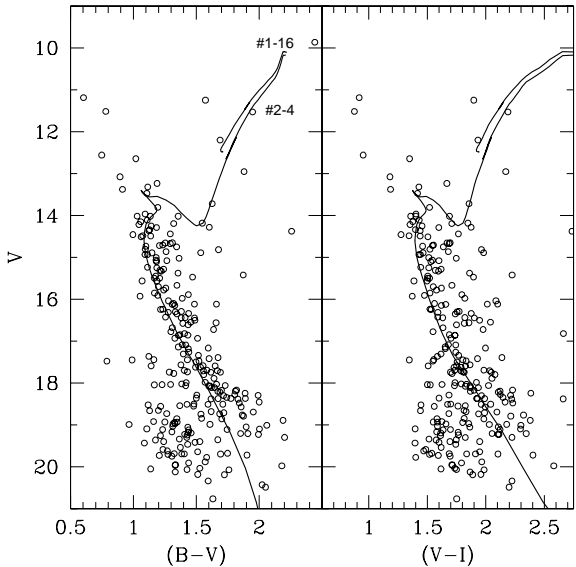,width=\columnwidth}} 
\caption{ {\bf Left panel}: NGC~2112 data in the $V$ vs.\ $B-V$ diagram,  
as compared to Girardi et al. \ (2000) isochrone of age 
$2.0\times10^9$ yr 
(solid line), for the metallicity $Z=0.012$. A distance 
modulus of $(m-M)_0=9.65$ mag, and a colour excess of E$(B-V)=0.63$ mag, 
have been derived. {\bf Right panel}: NGC~2112 data in the $V$ vs.\ $V-I$
diagram, as compared to Girardi et al. \ (2000) isochrone of age 
$2.0\times10^9$ yr 
(solid line), for the metallicity $Z=0.012$. A distance 
modulus of $(m-M)_0=9.65$ mag, and a colour excess of E$(V-I)=0.85$ mag, 
have been derived.} 
\end{figure} 
 
\subsection{Distance and age} 
As already mentioned, there is a considerable uncertainty
in the literature  
among different estimates of NGC~2112 distance and age,
as it is shown by the data reported in Table~2.
We have derived new estimates for these parameters as follows. 
First of all we have generated isochrones from Girardi et al. (2000)
by adopting the abundance determination by Brown et al.  (1996).
The value $[Fe/H]~=-0.15$ converts into the theoretical overall
metallicity Z=0.012.
Then, we have estimated the cluster age and distance simultaneously
by assuming that the reddening is that estimated in the previous
section : E$(B-V)$~=~0.63 mag.
The results are shown in Fig.~5. Here we consider the distribution of
the
stars in the $V$ vs $(B-V)$ plane (left panel)
and in the $V$ vs $(V-I)$ (right panel). Over-imposed 
in both diagrams is a 2.0 Gyrs
isochrone, which turns out to provide the best fit of the
overall CMD.
In doing the fit we paid attention to reproduce the TO,
the sub-giant region and the base of the RGB.
Noticeably, an important constraint is the star $\# 2-4$
(Richtler's numbering): this lies close to the theoretical
Red Giant Branch, and it is considered
a probable member (Brown et al. 1996).
On the other hand, the very red and bright star $\#1-16$
- considered a non-member (Brown et al. 1996) - clearly departs
from the fitting isochrone.\\
This fit has been reached by shifting the isochrone
vertically by $(m-M)~=~11.60 \pm 0.30$ mag, where the uncertainty
is here attributable to  both
the scarcely populated Red Giant region and the uncertainty 
in the reddening.\\
As a consequence,
the reddening corrected absolute distance
modulus becomes $(m-M)_o~=~9.65 \pm 0.30$ mag and  
the cluster distance from the Sun is $850 \pm 100$ pc.
While the distance we infer is basically 
consistent with previous suggestions, the age turns out to be
significantly younger. To give an estimate of the age uncertainty
we tried to fit the CMD with slightly younger and slightly
older isochrones, and we found that the best fit age was 
in the range 1.7 to 2.3 Gyrs.
\\
Finally, by assuming 8.5 kpc for the distance
of the Sun to the Galactic center,
we provide for NGC~2112 the following estimates
of the Galactocentric rectangular coordinates:
$X=9.2$, $Y=-0.36$ and $Z=-0.18$ kpc, which place NGC~2112
well within the Orion spiral arm.
We note however that the distance from the Galactic plane is larger than the 
scale height of the thin disk ($\approx 75$ pc). This situation is not
unusual, since there are several intermediate-age or old open clusters
presently at large distances from the Galactic plane: NGC~6791, 
NGC~2204 and  NGC~2420 for instance (see Friel 1995). 
Nevertheless they are  usually considered
members of the old thin disk which probably formed at fairly high $Z$
(Carraro \& Chiosi 1994, Friel 1995). 

\subsection{Consequences of a lower metal abundance}
In this section we re-derive estimates of NGC~2112 basic parameters
by assuming that its metal abundance is very low
($[Fe/H] \approx -1.0$), as found for instance by
Richtler (1985) and Richtler \& Kaluzny (1989).
The results are presented in Fig.~6.

\begin{figure} 
\centerline{\psfig{file=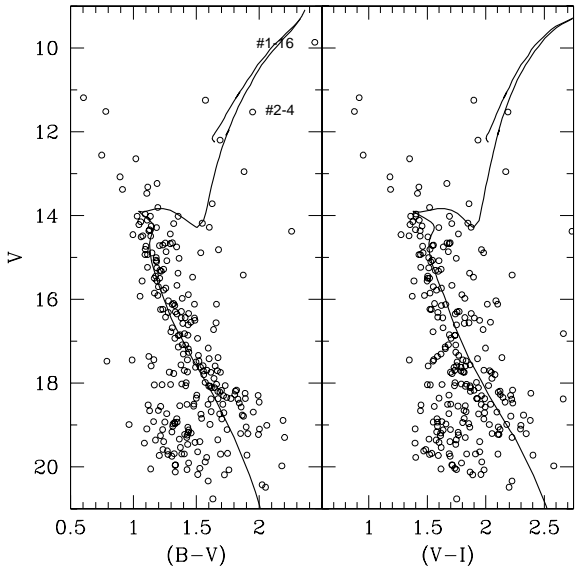,width=\columnwidth}} 
\caption{ {\bf Left panel}: NGC~2112 data in the $V$ vs.\ $B-V$ diagram,  
as compared to Girardi et al. \ (2000) isochrone of age 
$2.8\times10^9$ yr 
(solid line), for the metallicity $Z=0.003$. A distance 
modulus of $(m-M)_0=9.13$ mag, and a colour excess of E$(B-V)=0.82$ mag, 
have been obtained. {\bf Right panel}: NGC~2112 data in the $V$ vs.\ $V-I$
diagram, as compared to Girardi et al. \ (2000) isochrone of age 
$2.8\times10^9$ yr 
(solid line), for the metallicity $Z=0.003$. A distance 
modulus of $(m-M)_0=9.13$ mag, and a colour excess of E$(V-I)=1.07$ mag, 
have been obtained.} 
\end{figure} 

\noindent
In this case the theoretical overall abundance Z turns out
to be 0.003. By adopting this metallicity, it was not possible to 
derive a reasonable fit with the reddening obtained above.
On the contrary,
a good fit has been achieved with an isochrone of the age 2.8 Gyr ($15\%
$ uncertainty) yielding 
E$(B-V)= 0.82$ mag, and 
$(m-M)_0=9.13$ mag. Two comments are to be made.\\
The first one is that star $\#$2-4, usually considered a member
of NGC~2112, is clearly off the expected location of the RGB.
Secondly, the reddening implied by this metallicity is 
significantly larger than any previous estimate.
These facts seem to rule out the possibility that NGC~2112 is
very metal poor.

\section{NGC~2112 and the chemical evolution of the Galactic disk}
NGC~2112 is located toward the anticenter direction. As several
other clusters located in the Galactic sector between $l= 135
^{o}$
and $l = 225^{o}$ (Carraro et al.  2002), it is of intermediate-age
and has a lower than solar metal abundance ($[Fe/H] = -0.15$). 
It is interesting
to consider also this cluster in the age-metallicity
relationship (AMR) and in the Galactic abundance gradient, having
now at disposal updated estimates of the distance and age.
In fact this cluster has usually not been taken into account
in deriving global relationships for the Galactic Disk
(see Carraro et al.  1998). With an age of 2.0 Gyrs, a distance
from the Galactic center of 9.2 kpc
and a metal abundance of -0.15 dex, NGC~2112 perfectly fits
into both the age-metallicity and the distance-metallicity
relationships (see Carraro et al.  1998, Fig.~3).
However, if the metallicity were much lower, 
as derived in previous studies, the cluster
would place, in the AMR, close to Berkeley~21 (an
old open cluster for which a very low but 
uncertain metal abundance has been
reported, Friel \& Janes 1993, Hill \& Pasquini 1999, Tosi et al. 1998) ,
and it would hugely deviate from the mean Galactic abundance
gradient, concurring to steepen it.\\
It is quite clear that NGC~2112 -like other intermediate-age
metal poor clusters- plays an important role in shaping the
age-metallicity and age-distance
relationships  holding for the Galactic disk.
As a consequence, a 
proper motion study to better discriminate the physical members
and a more constrained metal
abundance determination are therefore really welcome.

   \begin{figure}
   \centerline{\psfig{file=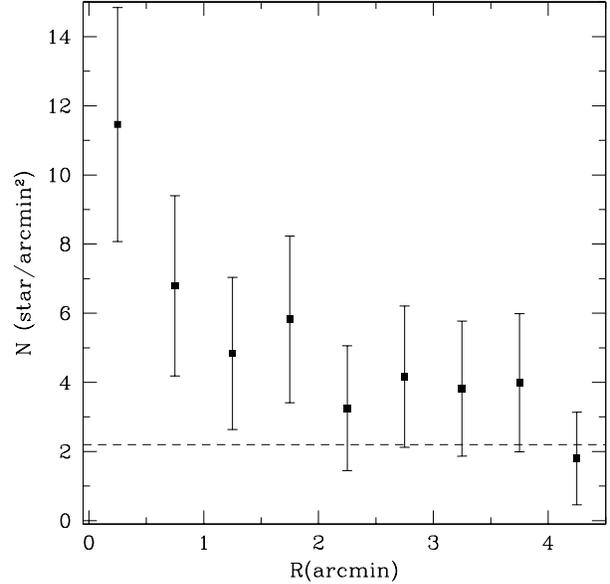,width=\columnwidth}} 
   \caption{ Star counts in the field of 
of NGC~2112 as a function of the radius. The dashed line is the
field number density estimate in the accompanying Galactic field.}
    \end{figure}

\section{Star counts and cluster size} 
We derive the surface stellar density by performing star counts
in concentric rings around stars $\#1-1$ (Richtler's 
numbering) selected as the approximate cluster center,
and then dividing by their
respective surfaces. The final density profile and the corresponding
Poisson error bars are shown in Fig.~7,
where we take into account all the measured stars brighter
than $V \approx 19$ mag.\\
The dashed line is the Galactic field number density derived 
from the 
accompanying field we observed in $V$ pass-band
centered $10^{\prime}$ eastward of NGC~2112. In this field we count 2.2 stars
per squared arcmin brighter than $V \approx 19$ mag.\\
The surface density decreases smoothly  over the whole region 
we covered, but the cluster starts to merge with the field already
at about 3.5-4.0 arcmin.
According to Lyng\aa~ (1987), NGC~2112 has a radius of about 9 arcmin.
This means that we were able to sample the cluster core (see Fig.~1),
where the contamination of field stars is minimized.
However it is very difficult to provide a precise
estimate of the cluster radius. In fact Friel \& Janes (1993)
identified two probable members 
($\#3-16$ and $\#3-17$, Richtler's numbering) 
outside the region we covered,
which probably belong to the cluster envelope.

\section{Conclusions} 
We have presented a new CCD $BVI$ photometric study of the intermediate-age 
open cluster NGC~2112. The CMDs we derive are much cleaner than
previous ones, and allow us to infer updated estimates of the cluster
basic parameters. In detail, we find that:
 
\begin{description} 
\item $\bullet$ the age of NGC~2112 is 2.0 Gyr, with a 15\,\% uncertainty; 
\item $\bullet$ the reddening $E_{B-V}$ turns out to be $0.63\pm0.14$ mag; 
\item $\bullet$ we place the cluster at about 0.85 kpc from the Sun toward 
the anti-center direction; 
\item $\bullet$ we show that Brown et al. (1996) estimate of the metallicity
is probably the most realistic one;
\item $\bullet$ combining together age, distance and metallicity, 
we suggest that this cluster is a genuine 
member of the old thin disk population. 
\end{description}

\noindent
As already noticed in the past, a proper motion study of NGC~2112
is really necessary to efficiently isolate cluster members from
non-members and 
then derive robust estimates of the cluster chemical abundance.

\section*{Acknowledgements} 
GC acknowledges the kind night assistance by Gigi Lessio.
This study has been financed by the Italian Ministry of 
University, Scientific Research and Technology (MURST) and the Italian 
Space Agency (ASI), and made use of Simbad and WEBDA databases.


\begin{thebibliography}{} 
\bibitem{} Brown J.A., Wallerstein G., Geisler D., Oke J.B., 1996, AJ
           112, 1551
\bibitem{} Carraro G., Ng Y.K., Portinari L, 1998, MNRAS 296, 1045
\bibitem{} Carraro G., Patat F., 1994, A\&A 317, 403
\bibitem{} Carraro G., Chiosi C., 1994, A\&A 288, 751
\bibitem{} Carraro G., Girardi L., Marigo P., 2002, MNRAS 332, 705
\bibitem{} Desidera S., Fantinel D., Giro E., Afosc User Manual, 2001
\bibitem{} Friel E.D., Janes K.A., 1993, A\&A 267, 75 
\bibitem{} Friel E.D., 1995, ARA\&A 33, 381 
\bibitem{} Geisler D., 1987, AJ 94, 84
\bibitem{} Geisler D., Clari\`a J.J. , Minniti D., 1991, AJ 102, 1836.
\bibitem{} Girardi L., Bressan A., Bertelli G., Chiosi C., 2000, 
           A\&AS 141, 371 
\bibitem{} Hill V., Pasquini L., 1999, A\&A 348, L21
\bibitem{} Landolt A.U., 1992, AJ 104, 340 
\bibitem{} Lyng\aa\ G., 1987, The Open Star Clusters Catalogue, 5th edition 
\bibitem{} Mermilliod J.-C., 1982, A\&A 109, 37
\bibitem{} Munari U., Carraro G., 1996, A\&A 314, 108 
\bibitem{} Patat F., Carraro G., 1995, A\&AS 114, 281 
\bibitem{} Patat F., Carraro G., 2001, MNRAS 325, 1591 
\bibitem{} Richtler T., 1985, A\&AS 59, 491
\bibitem{} Richtler T., Kaluzny J., 1989, A\&AS 81, 225
\bibitem{} Sharpless S., 1959, ApJS 4, 257
\bibitem{} Tosi M., Pulone L., Marconi G., Bragaglia A., 1998,
           MNRAS 299, 834
\bibitem{} Trumpler R.J., 1930, Lick Observ. Bull. 14, 154 
\end{thebibliography}
\end{document}